\input epsf
\newfam\scrfam
\batchmode\font\tenscr=rsfs10 \errorstopmode
\ifx\tenscr\nullfont
        \message{rsfs script font not available. Replacing with calligraphic.}
        
\else   
        \font\sevenscr=rsfs7
        \font\fivescr=rsfs5
        \skewchar\tenscr='177 \skewchar\sevenscr='177 \skewchar\fivescr='177
        \textfont\scrfam=\tenscr \scriptfont\scrfam=\sevenscr
        \scriptscriptfont\scrfam=\fivescr

\fi
\catcode`\@=11
\newfam\frakfam
\batchmode\font\tenfrak=eufm10 \errorstopmode
\ifx\tenfrak\nullfont
        \message{eufm font not available. Replacing with italic.}
        
\else
	
	\font\sevenfrak=eufm7 \font\fivefrak=eufm5
	\textfont\frakfam=\tenfrak
	\scriptfont\frakfam=\sevenfrak \scriptscriptfont\frakfam=\fivefrak
	
\fi
\catcode`\@=\active
\newfam\msbfam
\batchmode\font\twelvemsb=msbm10 scaled\magstep1 \errorstopmode
\ifx\twelvemsb\nullfont\def\Bbb{\bf}

	\message{Blackboard bold not available. Replacing with boldface.}
\else   \catcode`\@=11
        \font\tenmsb=msbm10 \font\sevenmsb=msbm7 \font\fivemsb=msbm5
        \textfont\msbfam=\tenmsb
        \scriptfont\msbfam=\sevenmsb \scriptscriptfont\msbfam=\fivemsb
        \def\Bbb{\relax\expandafter\Bbb@}
        \def\Bbb@#1{{\Bbb@@{#1}}}
        \def\Bbb@@#1{\fam\msbfam\relax#1}
        \catcode`\@=\active

\fi
        \font\eightrm=cmr8              \def\xrm{\eightrm}
        \font\eightbf=cmbx8             \def\xbf{\eightbf}
        \font\eightit=cmti10 at 8pt     \def\xit{\eightit}
        \font\eighttt=cmtt8             \def\xtt{\eighttt}
        \font\eightcp=cmcsc8
        \font\eighti=cmmi8              \def\xold{\eighti}
        \font\eightib=cmmib8             \def\xbold{\eightib}
        \font\teni=cmmi10               \def\old{\teni}
        \font\tencp=cmcsc10

        \font\twelvecp=cmcsc10 scaled\magstep1

	\font\eightsym=cmsy8

	 at10pt	
	\font\twelvehelvbold=phvb at12pt
	 at14pt
	\font\sixteenhelvbold=phvb at16pt

\def\noblackbox{\overfullrule=0pt}
\noblackbox

\newtoks\headtext
\headline={\ifnum\pageno=1\hfill\else
	\ifodd\pageno{\eightcp\the\headtext}{ }\dotfill{ }{\old\folio}
	\else{\old\folio}{ }\dotfill{ }{\eightcp\the\headtext}\fi
	\fi}
\def\makeheadline{\vbox to 0pt{\vss\noindent\the\headline\break
\hbox to\hsize{\hfill}}
        \vskip2\baselineskip}
\newcount\infootnote
\infootnote=0
\def\foot#1#2{\infootnote=1
\footnote{${}^{#1}$}{\vtop{\baselineskip=.75\baselineskip
\advance\hsize by -\parindent\noindent{\xrm #2}}}\infootnote=0$\,$}
\newcount\refcount
\refcount=1
\newwrite\refwrite
\def\oldsize{\ifnum\infootnote=1\xold\else\old\fi}
\def\ref#1#2{
	\def#1{{{\oldsize\the\refcount}}\ifnum\the\refcount=1\immediate\openout\refwrite=\jobname.refs\fi\immediate\write\refwrite{\item{[{\xold\the\refcount}]} 
	#2\hfill\par\vskip-2pt}\xdef#1{{\noexpand\oldsize\the\refcount}}\global\advance\refcount by 1}
	}
\def\refout{\catcode`\@=11
        \xrm\immediate\closeout\refwrite
        \vskip2\baselineskip
        {\noindent\twelvecp References}\hfill\vskip\baselineskip
        \baselineskip=.75\baselineskip
        \input\jobname.refs
        \baselineskip=4\baselineskip \divide\baselineskip by 3
        \catcode`\@=\active\rm}

\def\hepth#1{\href{http://xxx.lanl.gov/abs/hep-th/#1}{hep-th/{\xold#1}}}
\def\jhep#1#2#3#4{\href{http://jhep.sissa.it/stdsearch?paper=#2\%28#3\%29#4}{J. High Energy Phys. {\xbold #1#2} ({\xold#3}) {\xold#4}}}
\def\AP#1#2#3{Ann. Phys. {\xbold#1} ({\xold#2}) {\xold#3}}

\def\CMP#1#2#3{Commun. Math. Phys. {\xbold#1} ({\xold#2}) {\xold#3}}

\def\JHEP{\jhep}
\def\JMP#1#2#3{J. Math. Phys. {\xbold#1} ({\xold#2}) {\xold#3}}
\def\JPA#1#2#3{J. Phys. {\xbf A}{\xbold#1} ({\xold#2}) {\xold#3}}

\def\NPB#1#2#3{Nucl. Phys. {\xbf B}{\xbold#1} ({\xold#2}) {\xold#3}}
\def\NPPS#1#2#3{Nucl. Phys. Proc. Suppl. {\xbold#1} ({\xold#2}) {\xold#3}}
\def\PLB#1#2#3{Phys. Lett. {\xbf B}{\xbold#1} ({\xold#2}) {\xold#3}}

\def\PRD#1#2#3{Phys. Rev. {\xbf D}{\xbold#1} ({\xold#2}) {\xold#3}}

\newcount\sectioncount
\sectioncount=0
\def\section#1#2{\global\eqcount=0
	\global\subsectioncount=0
        \global\advance\sectioncount by 1
	\ifnum\sectioncount>1
	        \vskip2\baselineskip
	\fi
	\noindent
        \line{\twelvecp\the\sectioncount. #2\hfill}
		\vskip.8\baselineskip\noindent
        \xdef#1{{\old\the\sectioncount}}}
\newcount\subsectioncount
\def\subsection#1#2{\global\advance\subsectioncount by 1
	\vskip.8\baselineskip\noindent
	\line{\tencp\the\sectioncount.\the\subsectioncount. #2\hfill}
	\vskip.5\baselineskip\noindent
	\xdef#1{{\old\the\sectioncount}.{\old\the\subsectioncount}}}
\newcount\appendixcount
\appendixcount=0
\def\appendix#1{\global\eqcount=0
        \global\advance\appendixcount by 1
        \vskip2\baselineskip\noindent
        \ifnum\the\appendixcount=1
        \hbox{\twelvecp Appendix A: #1\hfill}\vskip\baselineskip\noindent\fi
    \ifnum\the\appendixcount=2
        \hbox{\twelvecp Appendix B: #1\hfill}\vskip\baselineskip\noindent\fi
    \ifnum\the\appendixcount=3
        \hbox{\twelvecp Appendix C: #1\hfill}\vskip\baselineskip\noindent\fi}
\def\acknowledgements{\vskip2\baselineskip\noindent
        \underbar{\it Acknowledgements:}\ }
\newcount\eqcount
\eqcount=0
\def\Eqn#1{\global\advance\eqcount by 1
\ifnum\the\sectioncount=0
	\xdef#1{{\old\the\eqcount}}
	\eqno({\oldstyle\the\eqcount})
\else
        \ifnum\the\appendixcount=0
	        \xdef#1{{\old\the\sectioncount}.{\old\the\eqcount}}
                \eqno({\oldstyle\the\sectioncount}.{\oldstyle\the\eqcount})\fi
        \ifnum\the\appendixcount=1
	        \xdef#1{{\oldstyle A}.{\old\the\eqcount}}
                \eqno({\oldstyle A}.{\oldstyle\the\eqcount})\fi
        \ifnum\the\appendixcount=2
	        \xdef#1{{\oldstyle B}.{\old\the\eqcount}}
                \eqno({\oldstyle B}.{\oldstyle\the\eqcount})\fi
        \ifnum\the\appendixcount=3
	        \xdef#1{{\oldstyle C}.{\old\the\eqcount}}
                \eqno({\oldstyle C}.{\oldstyle\the\eqcount})\fi
\fi}
\def\eqn{\global\advance\eqcount by 1
\ifnum\the\sectioncount=0
	\eqno({\oldstyle\the\eqcount})
\else
        \ifnum\the\appendixcount=0
                \eqno({\oldstyle\the\sectioncount}.{\oldstyle\the\eqcount})\fi
        \ifnum\the\appendixcount=1
                \eqno({\oldstyle A}.{\oldstyle\the\eqcount})\fi
        \ifnum\the\appendixcount=2
                \eqno({\oldstyle B}.{\oldstyle\the\eqcount})\fi
        \ifnum\the\appendixcount=3
                \eqno({\oldstyle C}.{\oldstyle\the\eqcount})\fi
\fi}
\def\multi{\global\advance\eqcount by 1}
\def\multieq#1#2{\xdef#1{{\old\the\eqcount#2}}
        \eqno{({\oldstyle\the\eqcount#2})}}
\newtoks\url
\def\Href#1#2{\catcode`\#=12\url={#1}\catcode`\#=\active#2}
\def\href#1#2{{#2}}
\def\hhref#1{{#1}}
\parskip=3.5pt plus .3pt minus .3pt
\baselineskip=14pt plus .1pt minus .05pt
\lineskip=.5pt plus .05pt minus .05pt
\lineskiplimit=.5pt
\abovedisplayskip=18pt plus 4pt minus 2pt
\belowdisplayskip=\abovedisplayskip
\hsize=14cm
\vsize=19cm
\hoffset=1.5cm
\voffset=1.8cm
\frenchspacing
\footline={}
\raggedbottom

\def\ss{\scriptstyle}
\def\sss{\scriptscriptstyle}
\def\*{\partial}
\def\punkt{\,\,.}
\def\komma{\,\,,}

\def\={\!=\!}
\def\small#1{{\hbox{$#1$}}}
\def\half{\small{1\over2}}
\def\fraction#1{\small{1\over#1}}

\def\tr{\hbox{\rm tr}}

\def\ie{{\tenit i.e.}}
\def\etal{{\tenit et al.}}

\def\a{\alpha}
\def\b{\beta}

\def\G{\Gamma}
\def\L{\Lambda}
\def\O{\Omega}

\def\C{{\Bbb C}}
\def\R{{\Bbb R}}
\def\H{{\Bbb H}}

\def\id{1\hskip-3.5pt 1}


\def\a{\alpha}
\def\ad{\dot\a}
\def\b{\beta}

\def\ld{\lambda^\dagger}

\def\G{\Gamma}
\def\L{\Lambda}
\def\Ld{{\L^\dagger}}

\def\R{{\Bbb R}}
\def\C{{\Bbb C}}
\def\H{{\Bbb H}}
\def\O{{\Bbb O}}
\def\K{{\Bbb K}}
\def\Kn{{\Bbb K}_\nu}


\ref\FradkinVasiliev{E. Fradkin and M.A. Vasiliev, {\xit ``On the
gravitational interaction of massless higher spin fields''},
\PLB{189}{1987}{89}.}

\ref\VasilievHS{M.A. Vasiliev, {\xit ``Equations of motion of
interacting massless fields of all spins as a free differential
algebra''},   
\PLB{209}{1988}{491};
{\xit ``Consistent equations for interacting massless fields of all
spins in the first order in curvatures''}, \AP{190}{1989}{59}; and
references therein.}

\ref\VasilievReview{M.A. Vasiliev, {\xit ``Progress in higher spin gauge
theories''}, \hepth{0104246}.}

\ref\VasilievVector{M.A. Vasiliev, {\xit ``Nonlinear equations for
symmetric massless higher spin fields in AdS${}_d$''},
\PLB{567}{2003}{139} [\hepth{0304049}].}

\ref\VasilievSp{M.A. Vasiliev, {\xit ``Relativity, causality,
locality, quantization and duality in the Sp(2M) invariant generalized
space-time''}, \hepth{0111119}; {\xit ``Higher spin theories and
Sp(2M)-invariant space-times''}, \hepth{0301235}.}

\ref\BandosOSp{I. Bandos, J. Lukierski, C. Preitschopf and
D. Sorokin, {\xit ``OSp supergroup manifolds, superparticles and
supertwistors''}, \PRD{61}{2000}{065009} [\hepth{9907113}].}

\ref\BandosFlat{I. Bandos, E. Ivanov, J. Lukierski and D. Sorokin,
{\xit ``On the superconformal flatness of AdS
Superspaces''}, \JHEP{02}{06}{2002}{040} [\hepth{0205104}].}

\ref\PlyushchayEtal{M. Plyushchay, D. Sorokin and M. Tsulaia, {\xit
``GL flatness of OSp(1$\ss\vert$2N) and higher spin field theory from
dynamics in tensorial spaces''}, in the proceedings of the
International Seminar on Supersymmetries and Quantum Symmetries SQS
'03, Dubna, July 2003, \hepth{0310297}.}

\ref\KalloshI{P. Claus, M. G\"unaydin, R. Kallosh, J. Rahmfeld and Y. Zunger,
{\xit ``Supertwistors as quarks of SU(2,2\hskip1pt{\eightsym\char'152}4)''},
\jhep{99}{05}{1999}{019} [\hepth{9905112}].}

\ref\KalloshII{P. Claus, R. Kallosh and J. Rahmfeld,
{\xit ``BRST quantization of a particle in AdS${}_{\sss5}$''},
\PLB{462}{1999}{285} [\hepth{9906195}].}

\ref\CederwallAdSTwistor{M. Cederwall, {\xit ``Geometric construction
of AdS twistors''}, \PLB{483}{2000}{257} [\hepth{0002216}].}

\ref\CederwallBengtsson{I. Bengtsson and M. Cederwall,
{\xit ``Particles, twistors and the division algebras''},
\NPB{302}{1988}{81}.}

\ref\CederwallOctonionic{M. Cederwall, 
{\xit ``Octonionic particles and the S${}^{\sss7}$ symmetry''},
\JMP{33}{1992}{388}.}

\ref\Sundborg{B. Sundborg, {\xit ``Stringy gravity, interacting
tensionless strings and massless higher
spins''}, \NPPS{102}{2001}{113} [\hepth{0103247}].}

\ref\Bonelli{G. Bonelli, {\xit ``On the tensionless limit of bosonic
strings, infinite symmetries and higher spins''}, \NPB{669}{2003}{159}
[\hepth{0305155}].}

\ref\BrinkCederwallPreitschopf{L. Brink, M. Cederwall and
C. Preitschopf,
{\xit ``N=8 superconformal algebra and the superstring''},
\PLB{311}{1993}{76} [\hepth{9303172}].}

\ref\CederwallPreitschopf{M. Cederwall and C. Preitschopf,
{\xit ``${\ss S}^{\sss7}$ and $\hat{{\ss S}^{\sss7}}$''}, 
\CMP{167}{1995}{373} [\hepth{9309030}].}

\ref\EnglertEtAl{F. Englert, A. Sevrin, W. Troost, A. Van Proeyen and 
P. Spindel, {\xit ``Loop algebras and superalgebras based on S(7)''}, 
\JMP{29}{1988}{281}.}

\ref\BandosIII{I. Bandos, P. Pasti, D. Sorokin and M. Tonin, {\xit 
``Superfield theories in tensorial superspaces and the dynamics of 
higher spin fields''}, \hepth{0407180}.}

\ref\Sudbery{A. Sudbery, 
{\xit ``Division algebras, (pseudo)orthogonal groups and spinors''}, 
\JPA{17}{1984}{939}.}

\ref\BerkovitsTen{N. Berkovits, 
{\xit ``A supertwistor description of the massless superparticle in
ten-dimensional superspace''}, 
\NPB{350}{1991}{193}.}

\ref\SorokinHigherSpinReview{D. Sorokin, {\xit ``Introduction to the
classical theory of higher spins''}, \hepth{0405069}.}

\ref\BouattaCompereSagnotti{N. Bouatta, G. Compere and A. Sagnotti,
{\xit ``An introduction to free higher-spin fields''}, Lectures given
at the Workshop on Higher Spin Gauge Theories, Brussels, May 2004,
\hepth{0409068}.}

\ref\Gunaydin{M. G\"unaydin, {\xit ``Unitary supermultiplets of
OSp(1$\ss\vert$32,R) and M-theory''}, \NPB{528}{1998}{432} [\hepth{9803138}].}

\ref\GunaydinTakemae{M. G\"unaydin and S. Takemae, {\xit ``Unitary
supermultiplets of OSp(8*$\ss\vert$4) and the AdS(7)/CFT(6) duality''}, 
\NPB{578}{2000}{405} [\hepth{9910110}].}

\ref\GunaydinMinic{M. G\"unaydin and Dj. Mini\'c, {\xit ``Singletons,
doubletons and M-theory''}, \NPB{523}{1998}{145} [\hepth{9802047}].}

\ref\SezginSundellI{E. Sezgin and P. Sundell, {\xit ``Doubletons and
5D higher spin gauge theory''}, \jhep{01}{09}{2001}{036} [\hepth{0105001}].}

\ref\SezginSundellII{E. Sezgin and P. Sundell, {\xit ``7D bosonic
higher spin gauge theory: Symmetry algebra and linearized
constraints''}, \NPB{634}{2002}{120} [\hepth{0112100}].}

\ref\SagnottiTsulaia{A. Sagnotti and M. Tsulaia, {\xit ``On higher
spins and the tensionless limit of string theory''},
\NPB{682}{2004}{83} [\hepth{0311257}].}

\ref\BeisertBianchiMoralesSamtleben{N. Beisert, M. Bianchi,
J.F. Morales and H. Samtleben, {\xit ``Higher spin symmetry and N=4
SYM''}, \jhep{04}{07}{2004}{058} [\hepth{0405057}].} 

\ref\SezginSundellHolography{E. Sezgin and P. Sundell, {\xit
``Massless higher spins and holography''}, \NPB{644}{2002}{303},
erratum \NPB{660}{2003}{403} [\hepth{0205131}].}


\headtext={Martin Cederwall: ``AdS Twistors for Higher Spin Theory''}

\line{
\epsfysize=1.7cm
\epsffile{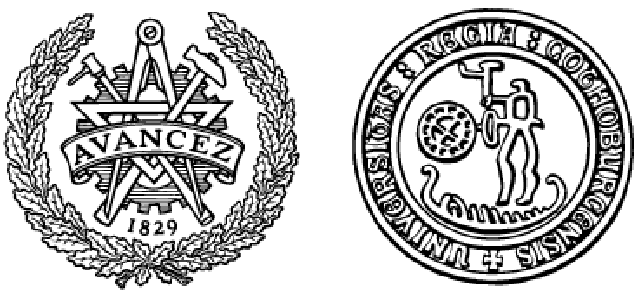}
\hfill}
\vskip-1.7cm
\line{\hfill G\"oteborg ITP preprint}
\line{\hfill hep-th/0412222}
\line{\hfill December, {\old2004}}
\line{\hrulefill}

\vfill

\centerline{\sixteenhelvbold AdS Twistors for Higher Spin Theory} 

\vskip1.6cm

\centerline{\twelvehelvbold Martin Cederwall}

\vskip.8cm

\centerline{\it Department of Theoretical Physics}
\centerline{\it G\"oteborg University and Chalmers University of Technology }
\centerline{\it S-412 96 G\"oteborg, Sweden}

\vskip1.6cm

{\narrower\noindent 
\underbar{Abstract:} We construct spectra of supersymmetric higher
spin theories in $D=4,5$ and 7 from twistors describing massless
(super-)particles on AdS spaces. A massless twistor transform is derived in a
geometric way from classical kinematics.
Relaxing the spin-shell constraints on twistor space gives an infinite tower
of massless states of a ``higher spin particle'', generalising
previous work of Bandos \etal.
 This can generically be done in a number of ways,
each defining the states of a distinct higher spin theory, and the
method provides a systematic way of finding these. We
reproduce known results in $D=4$, minimal supersymmetric 5- and
7-dimensional models, as well as supersymmetrisations of 
Vasiliev's Sp-models as special cases.
In the latter models a dimensional enhancement takes place,
meaning that the theory lives on a space of higher dimension than the
original AdS space, and becomes a theory of doubletons. This talk was
presented at the XIXth Max Born Symposium ``Fundamental Interactions
and Twistor-Like Methods'', September 2004, in
Wroc\l aw, Poland.
\smallskip}
\vfill

\vtop{\baselineskip=.8\baselineskip\xtt
\line{\hrulefill}
\catcode`\@=11
\line{martin.cederwall@fy.chalmers.se\hfill}
\catcode`\@=\active
\line{\hhref{http://fy.chalmers.se/tp/cederwall/}\hfill}
}

\eject

\section\Introduction{Introduction}It has been known for some time
that higher spin theory, \ie, theories of interacting massless higher
spin fields, should be formulated in anti-de Sitter space, or more
precisely, allow AdS space as a vacuum solution  
[\FradkinVasiliev]. The theory of higher spin fields was
subsequently developed in a series of papers [\VasilievHS]. 
 For excellent recent reviews, see
refs.
[\VasilievReview,\SorokinHigherSpinReview,\BouattaCompereSagnotti],
which also give an account to the earlier history of higher spin.

In recent years, there has been a growing interest in the theory of
massless higher spins. There is some hope that such a theory may provide a
geometric framework and a symmetry principle underlying string theory,
which in that case would be interpreted as a broken phase of higher
spin theory, where the higher spin fields have become massive. For the
possible connection between string theory and higher spin theory, see
refs. [\Sundborg,\SezginSundellHolography,\Bonelli,\SagnottiTsulaia,\BeisertBianchiMoralesSamtleben].

Higher spin theory has mostly been constructed using spinorial
oscillators. These are the kinds of models that will be discussed in
the present talk. Recently, constructions with vectorial variables,
for any dimension, have been performed [\VasilievVector], 
and we will have nothing to say
about these. The purpose of the work presented in this talk is to
present a unified framework for obtaining spinorial (twistorial)
variables for higher spin theory by relaxation of spin-shell
constraints for ordinary bosonic or supersymmetric particles. The
discussion is performed entirely at the kinematic, non-interacting, level.

\section\TwistorTransform{Twistor Transform for
Massless Particles on AdS}Consider AdS${}_{d+1}$ space with
radius $R$ as the hyperboloid
$$
x_Mx^M\equiv-(x^0)^2-(x^{0'})^2+\sum_{i=1}^d(x^i)^2=-R^2 \Eqn\AdSHyperboloid
$$
in flat space with signature $(2,d)$.
The trajectory of a massless particle, a light-like geodesic, 
is the intersection of the 
hyperboloid with a plane through the origin spanned by one light-like
 and one time-like vector, \ie, $x=\a X+\b P$, where $P^2=0$, 
$X^2<0$ and $X\cdot P=0$. $X$ may then be seen as the coordinate for
the location of the particle, fulfilling eq. (\AdSHyperboloid), and
$P$ as its momentum, being light-like and directed along the hyperboloid.
A plane is also defined by a bi-vector $\Pi^{MN}$ which is {\it simple}, 
meaning that it can be expressed in terms of two vectors as 
$\Pi^{MN}=X^{[M}P^{N]}$. 

\vskip2\parskip
\centerline{\epsffile{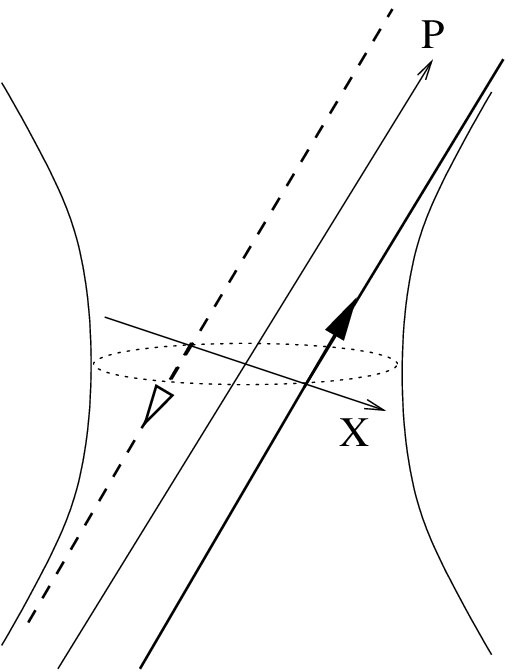}}
\centerline{\it Fig. 1. A light-like geodesic as the intersection of a
plane and a hyperboloid.}
\vskip2\parskip

The condition that $\Pi$ is simple 
can be expressed as 
$$
\Pi^{[MN}\Pi^{PQ]}=0\punkt\Eqn\Simplicity
$$ 
The properties of the vectors $X$ and $P$ imply in addition that
$$
\Pi_{MN}\Pi^{MN}=0\punkt\Eqn\Lightlikeness
$$

We want to find a twistor transform for a massless particle in AdS 
space. This means finding an expression $\Pi$ fulfilling
eqs. (\Simplicity) and (\Lightlikeness) as a spinor bilinear 
modulo some well defined constraints and transformations. The spinor 
should be a spinor under the AdS${}_{d+1}$ group Spin$(2,d)$. We restrict 
ourselves to the cases $d+1=4,5,7$, which are naturally related to 
the division algebras $\K_1=\R$, $\K_2=\C$ and $\K_4=\H$.
Using the isomorphisms $Sp(4;\Kn)\approx\hbox{Spin}(2,\nu+2)$, the 
spinors are 4-component with elements in $\Kn$.

The form of the twistor transform is
$$
\Pi^{MN}=\fraction8\Ld{}^I\G^{MN}\L^I\komma\Eqn\TheTwistorTransform
$$
where $I$ is an internal index that will turn out to run over two 
values, or equivalently, since the adjoint of $\hbox{Spin}(2,\nu+2)$ can be 
represented as a hermitean $4\times4$-matrix,
$\Pi^{A\dot B}=\half\L^{AI}\Ld{}^{I\dot B}$, or simply
$$
\Pi=\half\L\Ld\punkt\Eqn\AdSTwistor
$$
$\Pi$ contains the generator of AdS transformations, if $\L$ has 
canonical Poisson brackets with itself.

It is not possible to form a simple bivector from a single AdS 
spinor, the minimum number is two. That this is true can be checked 
explicitly for some simple specific choice of the plane. Take 
$\Pi^{\oplus0}\neq0$ (with the light-like $\oplus$ direction and the time-like
$0$ direction orthogonal) and the rest of the components vanishing. Using the 
gamma matrices of the Appendix, and denoting the 
spinor 
$$
\L=\left[\matrix{\lambda^\a\cr\mu_{\ad}\cr}\right]=\left[\matrix{k\cr 
l\cr m\cr n\cr}\right]\eqn
$$
(suppressing the index $I$), one needs
$$
\eqalign{
\Ld\G^{\oplus0}\L&\sim \bar kk+\bar ll\neq0\cr
\Ld\G^{\oplus,\nu+1}\L&\sim \bar kk-\bar ll=0\cr
\Ld\G^{\oplus i}\L&\sim \bar k\bar e_il+\bar le_ik=0\komma\quad 
i=1,\ldots\nu\cr
&\cr
\Ld\G^{\ominus0}\L&\sim\bar mm+\bar nn=0\cr
\Ld\G^{\ominus,\nu+1}\L&\sim\bar mm-\bar nn=0\cr
\Ld\G^{\ominus i}\L&\sim\bar m\bar e_in+\bar ne_im=0\komma\quad 
i=1,\ldots\nu\cr
}\eqn
$$
In order to satisfy the first three equations, two spinors are 
needed. Then the last three equations imply that $\mu=0$. The rest of 
the components of $\Pi$ mix $\lambda$ and $\mu$ and will thus vanish. 

A pair of spinors has an ``R-symmetry'', acting from the right on the 
twistor $\L\in\Kn^8$ of the form
$$
\L\sim\left[\matrix{
\cdot&\cdot\cr\cdot&\cdot\cr\cdot&\cdot\cr\cdot&\cdot\cr}\right]
\komma\eqn
$$
generated by anti-hermitean $2\times2$-matrices with entries in $\Kn$.
The corresponding groups are\foot\star{Here it becomes clear 
why AdS${}_{{\sss 11}}$ is not included, although one has the
isomorphism $\ss\hbox{Sp}(4;\O)\approx\hbox{Spin}(2,10)$. ``$\ss A_2(\O)$'' 
does not form a Lie
algebra, not even in the soft sense that $\ss A_1(\O)\approx S^7$ does
[\EnglertEtAl,\BerkovitsTen,\CederwallOctonionic,\BrinkCederwallPreitschopf,\CederwallPreitschopf]; 
it forms a
set of generators in a $\ss\hbox{Spin}(9)/G_2$ coset of $\ss\hbox{Spin}(9)$.}

$$
A_2(\Kn)=\left\{\matrix{\hbox{U}(1)\hfill\cr
\hbox{SU}(2)\times\hbox{U}(1)\hfill\cr
\hbox{Spin}(5)\hfill\cr}\right.\eqn
$$
Note that although these symmetries are ``internal'', and commute 
with the AdS generators, they are isomorphic to the groups of 
transverse rotations for a massless particle (except for the extra 
U(1), whose r\^ole is commented on below). 
Denote the corresponding generators $T$. The specific spinor $\L$ above 
clearly has $T=0$, since all generators contain ``$\mu\lambda$''. The rest 
of the conditions turn out to be consequences of the choice of frame. 
It is also clear that since $\Pi$ commutes with $T$, spinors should be 
counted modulo $T$-transformations. In fact, 
$$
T=\half\Ld\L\approx0\eqn
$$
is the gauge invariance of the twistor. The generators $T$ are
anti-hermitean due to the anti-hermiticity of the ``spinor metric''
$\epsilon$ (see the Appendix). The counting of physical 
degrees of freedom is straightforward:
$$
2\times4\nu-2(3\nu-2)=2\nu+4=2d\komma\eqn
$$
which is the correct number for the phase space of a massless particle 
in $d+1$ dimensions.

A twistor construction for massive particles was performed in refs. 
[\KalloshI,\KalloshII,\CederwallAdSTwistor]. It
was noted that four spinors were needed to form a plane whose intersection with
the hyperboloid is a massive geodesic for AdS${}_4$ and AdS${}_7$, while two
spinors were sufficient for AdS${}_5$.  
In the last of these cases, the U(1) generator was identified with the 
mass, and the SU(2) were gauge generators. In the 4- and 7-dimensional 
cases, it was necessary to break the ``R-symmetry'' $A_4(\Kn)$ by 
identifying a non-vanishing U(1) generator as the mass, which lead to 
a mixture of first and second class constraints. The present case of 
massless twistors is more uniform in the different dimensionalities, 
and only first class constraints are present.

We should stress a difference between twistor transforms on AdS 
space and Minkowski space. The division algebra construction 
[\CederwallBengtsson,\CederwallOctonionic] is 
natural in Minkowski space of dimensions $\nu+2=3$, 4, 6 and 10. The 
basic relation is 
$$
p\sim\lambda\ld\komma\Eqn\MinkowskiTwistor
$$
which due to Fierz identities in 
these dimensionalities ensure that $p$ is light-like. Such a 
construction is not possible in AdS, where the isometry group is 
semi-simple and does not contain translations. Any natural twistor 
transform will involve the whole twistor phase space, as in eq. 
(\AdSTwistor), not only a 
commuting half of it as in eq. (\MinkowskiTwistor). Also in Minkowski
space, however, eq. (\TwistorTransform) can be used as a definition
of the twistor transform, when the spinor $\lambda^\a$ is complemented
with its canonical conjugate $\mu_{\dot\a}$ and $\Pi$ is
interpreted as generators of conformal transformations. This
interpretation is relevant also in AdS space, where
eq. (\MinkowskiTwistor) is seen as the twistor transform on the boundary,
involving a commuting half of the AdS twistor, which will be utilised below. 

We finally note that the division algebra language is well suited to 
display certain Fierz identities relating the behavior under AdS 
transformations and under the ``R-symmetry''. We note that a 
$4\times4$ $\Kn$-valued matrix carrying two spinor indices $A$ and $\dot B$ 
decomposes into three
different irreducible representations: hermitean (the adjoint, and in
the complex case a singlet), 
anti-hermitean traceless (a 4-form, which in the real case is a
vector, in the complex case the adjoint and in the quaternionic case
self-dual) and the trace (singlet). 
If we form the twistor quadrilinear
$$
\Pi\Pi=\L\Ld\L\Ld\komma\Eqn\PiSquare
$$
it will be anti-hermitean and vanish due to $T=0$. The vanishing of the 4-form 
part shows that $\Pi$ is simple as in eq. (\Simplicity), and the 
singlet that $\Pi_{MN}\Pi^{MN}=0$. When we later relax the 
constraint $T=0$ to obtain ``higher spin particles'', such Fierz 
identities will relate Casimirs of the AdS group to Casimirs of 
R-symmetry.

The construction 
is straightforwardly extended to superparticles by introducing $2N$
anticommuting scalar variables $\Theta$ arranged in an $N\times2$
matrix, transforming from the left by $O(N;\Kn)$ and from the right by
$A_2(\Kn)$. The supertwistor
$$
\Xi=\left[\matrix{\Lambda\cr\Theta\cr}\right]\eqn
$$
transforms in the fundamental of the $N$-extended AdS${}_{\nu+3}$
supergroup $OSp(N|4;\Kn)$. The constraint structure is not affected,
we now have
$T=\half\Xi^\dagger\Xi
=\half(\Lambda^\dagger\Lambda+\Theta^\dagger\Theta)\approx0$, still
generating $A_2(\Kn)$.

\section\HigherSpinTwistors{Twistors for Higher Spin}We will now relax 
the constraint $T=0$ in order to incorporate higher spin massless 
states in the model. Since four spinors are needed in the 4- and 
7-dimensional models to describe massive states, it is essentially 
clear that such states will not appear. In a twistor model, the 
on-shell constraint is not a mass-shell constraint, but rather a 
spin-shell constraint, so relaxing it will lead to a multitude of 
spins. In the 5-dimensional case we need to be careful---since the 
U(1) generator measures the mass we are only to relax the SU(2) 
constraint (which matches nicely with the observation that this is 
isomorphic to transverse rotations for a massless particle in five 
dimensions).

It is possible to relax only part
of the spin-shell constraints. In AdS${}_4$ with gauge group U(1),
there is no choice, but in the other cases keeping as gauge group 
any subgroup of SU(2) or
Spin(5) (smaller than the group itself) should define a distinct
higher spin theory. Since the
generators $T$ are AdS scalars, this procedure is covariant, and will
result in restrictions on the massless representations occurring. In
this sense, the ``smallest'' higher spin theory should be given by the
``biggest'' subgroup. For example, one could consider a higher spin
theory on AdS${}_5$ with internal ``spin manifold'' (the non-gauged part
of the R-symmetry)
$\hbox{SU}(2)/U(1)=S^2$ or a theory on AdS${}_7$ with spin manifold
$Spin(5)/(\hbox{SU}(2)\times\hbox{SU}(2))$. These choices give the
minimal models considered by Sezgin and Sundell
[\SezginSundellI,\SezginSundellII]. 
If, on the other hand, the subgroup chosen is the
trivial one, one obtains the ``Sp space-times'' of
ref. [\VasilievSp]. 

Quantisation is straightforward and goes as follows. States in unitary
representations of the AdS group are formed by letting an
(anti-)commuting subset of variables in $\Xi$ (a configuration space)
act on a (preferably scalar) vacuum. 
States are thus obtained as polynomials in half of the supertwistor variables.
For an ordinary superparticle,
where $T\approx0$, the constraints are implemented by only considering
states that are R-symmetry singlets. If a subset of generators are
kept as gauge generators, as in the minimal models, they are treated
accordingly. A natural choice of configuration space for the bosonic
variables is the upper half $\lambda$ of $\Lambda$, which is a pair of spinors
under the Lorentz group acting on the boundary of AdS space. 

The minimal models in 4, 5 and 7 dimensions have gauge groups $\emptyset$,
U(1)$\times$U(1) and SU(2)$\times$SU(2), respectively. The two factors
act each on one column of $\Xi$ and do not mix them---the two spinors
decouple with this choice of spin constraints. The states of the
minimal theories are thus obtained as the tensor product of
two representations obtained from a single AdS spinorial oscillator,
\ie, as the tensor product of two singletons/doubletons, as in
refs. [\SezginSundellI,\SezginSundellII].  

Let us for a moment dwell on the ``maximal'' models, the
supersymmetric versions of the Sp-models.
It is trivial to write an action for the higher spin particle,
$$
S=\int d\tau\Xi^\dagger\dot\Xi\punkt\eqn
$$
(for AdS${}_5$ we should still include a Lagrange multiplier for the 
mass U(1)). From just inspecting the action, 
we see that the higher spin particle 
is invariant under a much larger symmetry than the AdS group. In 4 
and 7 dimensions, where there are no constraints, this is the real
orthosymplectic supergroup acting on the number of real fermionic and
bosonic components of the 
supertwistor, \ie, OSp(2$N\vert$8) (as noted in ref. [\BandosOSp]) and
OSp(8$N\vert$32),  
respectively. In AdS${}_5$ we have the subgroup of OSp(4$N\vert$16) that 
commutes with U(1), which is SU(2$N\vert$4,4). 
The corresponding groups acting on the (bosonic part of)
configuration space, and thus 
extending the Lorentz symmetry on the boundary, are SL(4), SL(4,$\C$)
and SL(16). 

As mentioned earlier, the twistor parametrisation of the AdS generators $\Pi$
imply a direct relation between Casimirs of the AdS group and of the internal
symmetry. This relation follows immediately from equations like (\PiSquare) and
reads 
$$
\tr\Pi^n=\tr T^n\punkt\eqn
$$
One might worry that once the spin-shell constraint is relaxed, the
relation to space-time is lost. The following consideration shows that
this is not the case.
A simple bi-vector, describing a plane, can be written as 
$\Pi^{MN}=X^{[M}P^{N]}$, where $X^2=-R^2$, $X\cdot P=0$ ($P$ is tangent to the
hyperboloid) and $P^2=0$. When
$T\neq0$, the best one can do is $\Pi^{MN}=X^{[M}P^{N]}+S^{MN}$. In order to
still have $\tr\Pi^2=\tr T^2$, one should demand that $X_MS^{MN}=0$
(spin is in the AdS tangent space) and 
$P_MS^{MN}=0$ (spin is transverse). 
In addition, $S$ is defined modulo $\delta S^{MN}=V^{[M}P^{N]}$,
with $V_MX^M=V_MP^M=0$, which can be absorbed in a redefinition of $X$ (a
reflection of gauge invariance). Such an $S$ is restricted to lie in the
transverse rotations, and one will have $\tr\Pi^n=\tr S^n$. This also
shows that although $T$ are the generators of an algebra of {\it
internal} rotations, the parametrisation of the AdS generators in
terms of spinors (the twistor transform) implies the identification of
all spin quantum numbers with quantum numbers with respect to $T$, and
the isomorphism between the internal algebra and the algebra of
transverse spin degrees of freedom becomes a physical identification.

In the spirit of ref. [\VasilievSp], we can ask questions about the causal
structure in such a theory. Vasiliev showed in ref. [\VasilievSp] that
the causal structure is determined by the maximal Clifford
algebras contained in the representations of the bilinears in $\lambda$.
They must contain matrices acting on 4-component real, 4-component
complex and 16-component real spinors, respectively, considering that
the subalgebras acting linearly on the configuration space of boundary
spinors are 
SL(4), SL(4,$\C$) and SL(16). 
We note that these groups are large enough to contain as
subgroups the Lorentz groups in higher dimensions, and in fact, since
we consider pairs of division algebra spinors, the dimensions
corresponding to choosing the next larger division algebra. This means
that the boundary in the maximal higher spin model in dimensions 4, 5
and 7 may be considered as 4-, 6- and 10-dimensional. Interpreted in
the higher dimensionalities, the pair of spinors becomes a single
spinor, and the corresponding states are those of a doubleton in 5, 7
and 11 dimensions.

The maximal
dimensionality of the Minkowski space is then 4, 6 and 10,
respectively, where the boundary spinors are 2-component complex,
quaternionic and octonionic.
When we thus consider the 
$\hbox{SL}(2,\K_{2\nu})\approx\hbox{Spin}(1,2\nu+1)$ subgroup of the
boundary group, the bilinears in $\lambda$ decompose into a vector and a set of
tensorial charges. In the theory originating from AdS${}_4$, where the
boundary has now become 4-dimensional, we have the vectorial coordinates and a
2-form (this is the ``tensorial'' model considered by Bandos
\etal. [\BandosOSp,\BandosFlat,\BandosIII,\PlyushchayEtal]). 
In the theory on AdS${}_5$,
where the boundary is enhanced to be 6-dimensional, we have the vector
and an SU(2)-triplet of self-dual 3-forms, and in the AdS${}_7$ model, with
a 10-dimensional boundary, there is the vector and a self-dual 5-form.

Starting out from a theory in AdS${}_{d+1}$ space, with $d=4,5,7$, we
end up with a different space-time interpretation, where the
``coordinates'' are the ordinary coordinates together with a set of
forms, ``central charge coordinates''. The latter provide an
alternative picture of the spin degrees of freedom in higher spin
theory. However, the general picture is more intricate than in 4
dimensions, where the alternative space-time also is
4-dimensional. We see that the alternative descriptions of the
5- and 7-dimensional AdS theories are theories in 6- and 
10-dimensional Minkowski space. 

Some more things can be made explicit about the relation between the
two descriptions/interpretations. Let us take the AdS${}_7$ theory as
an example. The original AdS group is
$\hbox{Spin}(2,6)\approx\hbox{Sp(4;$\H$)}$, and the internal symmetry
group is $\hbox{Spin}(5)\approx A_2(\H)$. $\hbox{Sp}(4;\H)\times A_2(\H)$ is
a subgroup of the group of all symplectic transformations on the
bispinor, which is Sp(32). Restricting to transformations on
configuration space, \ie, on half the spinor, breaks 
$\hbox{Sp(4;\H)}\times A_2(\H)$ to 
$\hbox{SL(2;\H)}\times
A_2(\H)\approx\hbox{Spin}(1,5)\times\hbox{Spin}(5)$ and Sp(32) to
SL(16). When we looked for the maximal Clifford algebra above, that
procedure singled out $\hbox{Spin}(1,9)\approx\hbox{SL}(2;\O)$. 
The Spin(1,5) group is smaller than the Lorentz group on the tangent
space of AdS${}_7$. It can be identified with the Lorentz group on
the conformally Minkowski boundary of AdS${}_7$. This gives a picture
of what happens: the boundary Lorentz group is a common subgroup of
the AdS group and the boundary symmetry group  (in this case Sp(16)). While
adjoining a radial coordinate gives the AdS picture, the alternative
picture is obtained by supplementing the boundary coordinates with a
number of ``spin degrees of freedom'', manifested as tensorial
variables. While this discussion refers to the configuration space, or
the boundary variables, one may equally well consider the
dimensionally enhanced model as a bulk theory on AdS${}_{11}$. There
the states are those obtained from a single supertwistor, which means
that it is a doubleton in the sense of ref. [\Gunaydin,\GunaydinMinic].

\section\Conclusions{Concluding Remarks}We have given a systematic way
of deriving different versions of bosonic and supersymmetric higher
spin theory on anti-de Sitter spaces of dimension 4, 5 and 7, \ie, the
dimensionalities related to the real, complex and quaternionic
division algebras. The construction, as it stands is first-quantised,
and can be understood as describing the dynamics of a (free) higher
spin particle. 

The main message is that starting with a twistorial description of
particle mechanics, the spin-shell constraints may be relaxed in a
systematic and controlled way to yield higher spin degrees of freedom.
Using this prescription produces all known spinorial descriptions of
higher spin kinematics as special cases. So far, nothing really new has
come out of the present work, although we think it provides a nice and
unified framework for known models, and a better understanding of
their respective r\^oles and relations.

It would be
interesting, especially considering the potential relation of higher
spin theory to string theory
[\Sundborg,\SezginSundellHolography,\Bonelli,\SagnottiTsulaia,\BeisertBianchiMoralesSamtleben],
to see if one can use the particle action and couple it
to background fields describing exactly the states produced by the
first-quantised action itself, and what information about the interacting
theory may be obtained this way.

\appendix{Spinors and Gamma Matrices}Here, we set the conventions
for the gamma matrices used in the twistor transform 
(\TheTwistorTransform). They can be given in a unified notation for the
three cases. We denote by $e_i$, $i=1,\ldots,\nu$, the standard 
orthonormal basis for the division algebra $\K_\nu$.
For the flat embedding space with signature $(2,\nu+2)$ we use 
light-cone coordinates $M=(\oplus,\ominus,\mu)=(\oplus,\ominus,+,-,i)$ 
and scalar product
$V\cdot W=-V^\oplus W^\ominus-V^\ominus W^\oplus-V^+W^--V^-W^++V^iW^i$.

A spinor under the AdS group belongs to the fundamental representation
of Sp$(4;\K_\nu)$, \ie, it is a 4-component column with entries in 
$\K_\nu$. We use a dotted/undotted notation for spinors, and in addition primed
and unprimed spinor indices (since there generically are two 
chiralities). 
The two spinor representations, with indices $A$ and $A'$, both
decompose into as ${}^A\rightarrow({}^\a,{}_{\dot\a})\leftarrow{}^{A'}$
under the subgroup SL(2;$\K_\nu$) $\approx$ Spin(1,$\,\nu+1$).
The gamma matrices (or, strictly speaking, sigma 
matrices) acting on one chirality (the unprimed one that is chosen
for the twistors) are
$$
\eqalign{
&\G^{\oplus A'}{}_B=
\left[\matrix{\sqrt2\,\id^\alpha{}_\beta&0\cr0&0\cr}\right]
\komma\quad
\G^{\ominus A'}{}_B=
\left[\matrix{0&0\cr0&\sqrt2\,\id_{\dot\alpha}{}^{\dot\beta}\cr}\right]
\komma\quad   \cr
&\G^{\mu A'}{}_B=
\left[\matrix{0&\tilde\gamma^{\mu\alpha\dot\beta}\cr
\gamma^{\mu}{}_{\dot\alpha\beta}&0\cr}\right]
\komma   \cr}\eqn
$$
and on the other one
$$
\eqalign{
&\tilde\G^{\oplus A}{}_{B'}=
\left[\matrix{0&0\cr0&-\sqrt2\,\id_{\dot\alpha}{}^{\dot\beta}\cr}\right]
\komma\quad
\tilde\G^{\ominus A}{}_{B'}=
\left[\matrix{-\sqrt2\,\id^\alpha{}_\beta&0\cr0&0\cr}\right]
\komma\quad  \cr
&\tilde\G^{\mu A}{}_{B'}=
\left[\matrix{0&\tilde\gamma^{\mu\alpha\dot\beta}\cr
\gamma^{\mu}{}_{\dot\alpha\beta}&0\cr}\right]
\komma  \cr}\eqn
$$
where $\gamma^\mu$, $\tilde\gamma^\mu$ are 
SL(2;$\K_\nu$) $\approx$ Spin(1,$\,\nu+1$) gamma 
matrices:
$$
\eqalign{
&\gamma^+{}_{\dot\alpha\beta}=\left[\matrix{\sqrt2&0\cr0&0\cr}\right]
\komma\quad
\gamma^-{}_{\dot\alpha\beta}=\left[\matrix{0&0\cr0&\sqrt2\cr}\right]
\komma\quad
\gamma^i{}_{\dot\alpha\beta}=\left[\matrix{0&\bar e_i\cr e_i&0\cr}\right]
\cr
&\tilde\gamma^{+\alpha\dot\beta}=\left[\matrix{0&0\cr0&-\sqrt2\cr}\right]
\komma\quad
\tilde\gamma^{-\alpha\dot\beta}=\left[\matrix{-\sqrt2&0\cr0&0\cr}\right]
\komma\quad
\tilde\gamma^{i\alpha\dot\beta}=\left[\matrix{0&\bar e_i\cr e_i&0\cr}\right]
\punkt\cr}\eqn
$$
The matrices $\Gamma^{MN}$ used in the construction of the plane 
defining the geodesics are constructed as
$$
\Gamma^{MNA}{}_B=
\half\bigl(\tilde\Gamma^M\Gamma^N-\tilde\Gamma^N\Gamma^M\bigr)^A{}_B
\komma\eqn
$$
and the twistor bilinear is given by 
$$
\Pi^{MN}=\half\Ld\G^{MN}\L\equiv\half\L^\dagger_A\Gamma^{MNA}{}_B\L^B
\equiv\half\L^{\dagger\dot A}\varepsilon_{\dot AB}\Gamma^{MNB}{}_C\L^C
\komma\eqn
$$
where the anti-hermitean ``spinor metric''
$$
\varepsilon_{\dot AB}=
\left[\matrix{0&\id_{\dot\alpha}{}^{\dot\beta}\cr
-\id^\alpha{}_\beta&0\cr}\right]
\eqn
$$ 
is used to lower the spinor index, and
where $\dagger$ implies division algebra hermitean conjugation.
In the real case, dots are of course superfluous, and there is only 
one chirality. The above formul\ae\ are still correct, and primed and
unprimed indices are then related via
$$
E^{A'}{}_B=\left[\matrix{0&\varepsilon^{\alpha\beta}\cr
\varepsilon_{\alpha\beta}&0\cr}\right]
\punkt\eqn
$$

\vfill\eject

\acknowledgements
The author is grateful to Jerzy Lukierski, Per Sundell, Dmitri
Sorokin and, in particular, to Michail Vasiliev for discussions and
explanations of their work, and to Murat G\"unaydin for comments. He
would also like to thank the organizers of the XIXth Max Born
Symposium in Wroc\l aw, September 2004, 
where this work was presented, as well as the
organisers of the 9th Adriatic Meeting, Dubrovnik, September 2003,
where most of the present work was done, for their hospitality.

\refout

\end